\input epsf.sty
\nopagenumbers

\font\bigbf=cmbx10 scaled\magstep1

\line{\hss\bigbf LIGHT PROPAGATION IN INHOMOGENEOUS UNIVERSE \hss}

\bigskip

\line{\hss
Premana W. Premadi,\footnote{$\,^1$}{Institute of Astronomy, Tohoku University
                                  Sendai 981, Japan}
Hugo Martel,\footnote{$\,^2$}{Department of Astronomy, University of Texas
	                  Austin, TX 78712, USA}
and
Richard A. Matzner\footnote{$\,^3$}{Center for Relativity, University of Texas
	                          Austin, TX 78712, USA}$^,$
   \footnote{$\!^4$}{Department of Physics, University of Texas
	                          Austin, TX 78712, USA}\hss}

\bigskip\smallskip

\noindent{\bf 1.\quad Overview}

\medskip

Using a multi-plane lensing method that we have
developed (Premadi 1996;
Premadi, Martel, \& Matzner 1998), we follow the evolution
of light beams as they propagate through
inhomogeneous universes. We use a P$^3$M 
code to simulate the formation and evolution of large-scale structure.
The resolution of the simulations is increased to
sub-Megaparsec scales by using a Monte Carlo method to locate galaxies
inside the computational volume according to the underlying particle
distribution. The galaxies are approximated by isothermal spheres, with
each morphological type having its own distribution of masses
and core radii (Jarosz\'ynski 1992). The morphological types are chosen in
order to reproduce the observed morphology-density relation (Dressler 1980).
This algorithm has an effective resolution of
9 orders of magnitudes in length, from the size of superclusters down to the
core radii of the smallest galaxies.

We consider cold dark matter models normalized to {\it COBE}, 
and perform a large parameter survey by varying the cosmological parameters
$\Omega_0$, $\lambda_0$, $H_0$, and $n$ (the tilt of the primordial power
spectrum). The values of $n$ are chosen by imposing particular values or
$\sigma_8$, 
the rms mass fluctuation at a scale of $8h^{-1}\rm Mpc$. We use the
power spectrum given in Bunn \& White (1997). 
Table 1 gives the values of the parameters for all models. This is
the largest parameter survey ever done is this field.

\bigskip

\line{\hss Table 1. Parameters for the various models \hss}
\bigskip

\line{\hss
\vbox{
\halign{
\strut\hfil#\hfil\quad&\hfil#\hfil\quad&\hfil#\hfil\quad&\hfil#\hfil\quad&
\hfil#\hfil\qquad\qquad&\hfil#\hfil\quad&\hfil#\hfil\quad&\hfil#\hfil\quad&
\hfil#\hfil\quad&\hfil#\hfil\cr
\noalign{\hrule}
$\Omega_0$ & $\lambda_0$ & $H_0$ & $n$ & $\sigma_8$ &
$\Omega_0$ & $\lambda_0$ & $H_0$ & $n$ & $\sigma_8$ \cr
\noalign{\hrule}
1.0 & 0.0 & 65.0 & 0.7234 & 0.9 & 0.2 & 0.0 & 65.0 & 1.3188 & 0.5 \cr
1.0 & 0.0 & 55.0 & 0.8465 & 1.0 & 0.2 & 0.0 & 75.0 & 1.2190 & 0.5 \cr
1.0 & 0.0 & 65.0 & 0.7698 & 1.0 & 0.2 & 0.0 & 75.0 & 1.2979 & 0.6 \cr
1.0 & 0.0 & 75.0 & 0.7094 & 1.0 & 0.2 & 0.0 & 75.0 & 1.3648 & 0.7 \cr
1.0 & 0.0 & 85.0 & 0.6605 & 1.0 & 0.7 & 0.3 & 65.0 & 0.7720 & 0.9 \cr
1.0 & 0.0 & 65.0 & 0.8120 & 1.1 & 0.7 & 0.3 & 75.0 & 0.7042 & 0.9 \cr
1.0 & 0.0 & 65.0 & 0.8506 & 1.2 & 0.7 & 0.3 & 65.0 & 0.8601 & 1.1 \cr
1.0 & 0.0 & 75.0 & 0.7893 & 1.2 & 0.7 & 0.3 & 75.0 & 0.7912 & 1.1 \cr
1.0 & 0.0 & 65.0 & 0.8861 & 1.3 & 0.5 & 0.5 & 65.0 & 0.7808 & 0.8 \cr
0.7 & 0.0 & 65.0 & 0.8461 & 0.9 & 0.5 & 0.5 & 75.0 & 0.7049 & 0.8 \cr
0.7 & 0.0 & 75.0 & 0.7773 & 0.9 & 0.5 & 0.5 & 65.0 & 0.8807 & 1.0 \cr
0.7 & 0.0 & 65.0 & 0.9346 & 1.1 & 0.5 & 0.5 & 75.0 & 0.8024 & 1.0 \cr
0.7 & 0.0 & 75.0 & 0.8648 & 1.1 & 0.2 & 0.8 & 65.0 & 0.9326 & 0.6 \cr
0.5 & 0.0 & 65.0 & 0.9457 & 0.8 & 0.2 & 0.8 & 75.0 & 0.8273 & 0.6 \cr
0.5 & 0.0 & 75.0 & 0.8686 & 0.8 & 0.2 & 0.8 & 65.0 & 1.0702 & 0.7 \cr
0.5 & 0.0 & 65.0 & 1.0439 & 1.0 & 0.2 & 0.8 & 55.0 & 1.2057 & 0.8 \cr
0.5 & 0.0 & 75.0 & 0.9656 & 1.0 & 0.2 & 0.8 & 65.0 & 1.0702 & 0.8 \cr
0.2 & 0.0 & 55.0 & 1.2187 & 0.3 & 0.2 & 0.8 & 75.0 & 0.9629 & 0.8 \cr
0.2 & 0.0 & 65.0 & 1.0966 & 0.3 & 0.2 & 0.8 & 85.0 & 0.8749 & 0.8 \cr
0.2 & 0.0 & 75.0 & 0.9993 & 0.3 & 0.2 & 0.8 & 65.0 & 1.1269 & 0.9 \cr
0.2 & 0.0 & 85.0 & 0.9191 & 0.3 & 0.2 & 0.8 & 65.0 & 1.1568 & 1.0 \cr
0.2 & 0.0 & 75.0 & 1.1228 & 0.4 \cr
\noalign{\hrule}
}
}\hss}

\vfill\eject

\noindent{\bf 2.\quad The Ray Shooting Experiments}

\medskip

For each model, we perform numerous ray-tracing experiments, 
propagating beams of $341^2$ light rays back in time
up to redshifts $z=3$ or 5. Distance between two neighboring rays
is about 1 arcsecond. This work is still in progress.
So far, we have performed between 20 and 40 experiments for each
model. Even with these
small numbers of runs, we can already see significant differences 
between the various cosmological models.
Figure 1 shows the individual contribution of each lens plane to
the shear, as a function of the lens redshift, for sources located
at $z=3$ and $z=5$, and for 3 cosmological models:
An Einstein-de~Sitter model, an open model with $\Omega_0=0.2$ and
$\lambda_0=0$, and a flat model with $\Omega_0=0.2$ and $\lambda_0=0.8$
(these results are obtained by averaging over many experiments).
The lenses planes contributing the most are located at intermediate
redshifts, in spite of the fact that structures are more evolved at low
redshifts, and galaxies are more crowded at high redshift. The shear
peaks at lower redshift for the Einstein-de~Sitter model than for the
other two models. We interpret this effect as a consequence of the freeze-out
of the growth of perturbations in models with $\Omega_0<1$. In such models,
structures present at low redshift must have formed earlier than in the 
Einstein-de~Sitter model. Hence, such models tend to have more structure at
intermediate redshift for  a given value of $\sigma_8$.

\vskip-1cm
\epsfxsize=13.0cm
\epsfbox{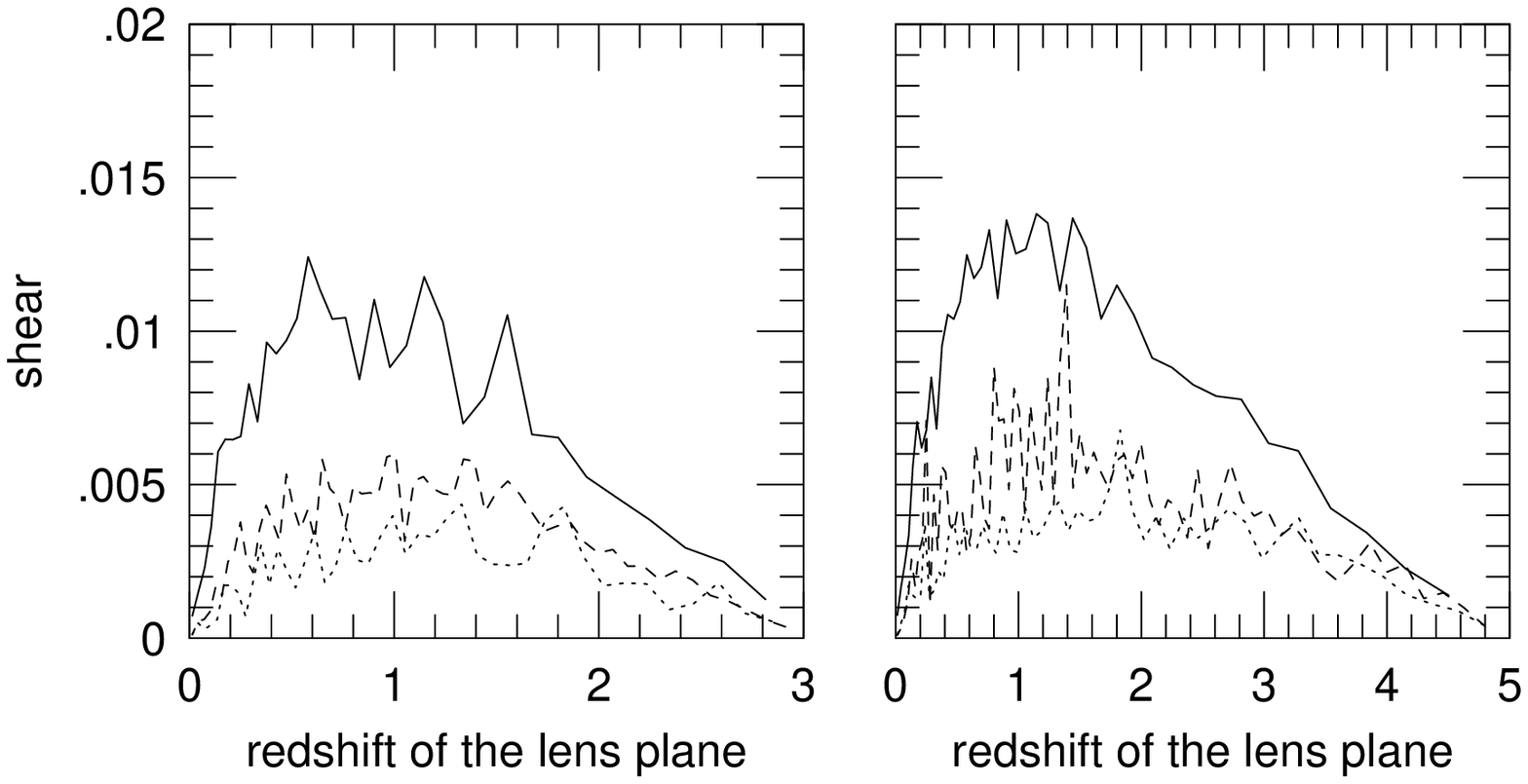}
\vskip-8.5cm
\noindent Fig 1. Individual contribution of each lens plane to the shear
vs redshift of the lens plane, for sources located at $z=3$ (left panel)
and $z=5$ (right model). Solid curves: Einstein-de~Sitter model;
dotted curves: open model; dashed curves: flat, cosmological constant model.

\bigskip

Using these experiments, we can locate actual sources on the source plane,
and compute their images on the image plane.
In Figure 2, we plot the images of distant
circular sources, to illustrate interesting cases that occurred in some
experiments: magnification and shear (Fig. 2a), double image (Fig. 2b),
Einstein ring (Fig. 2c), and triple image (Fig. 2d).

\bigskip\smallskip

\noindent{\bf 3.\quad Preliminary Results}

\medskip

We can use the properties of the images to compute lensing statistics that
can eventually be compared with observations. This is the ultimate goal of 
this project, and it will require many more experiments than the ones we
have performed so far. For illustrative purposes, we computed, for
various models, the probability that a lensed quasar will have multiple images.
The results are presented in Table 2. As we see, the probabilities are small.
Hence, most lensed quasars are simply magnified, with no image splitting.
The results appear to vary among different cosmological models, but the
number of runs is still to small to be statistically significant.

\vskip-3cm
\epsfxsize=13cm
\hskip0.9cm\epsfbox{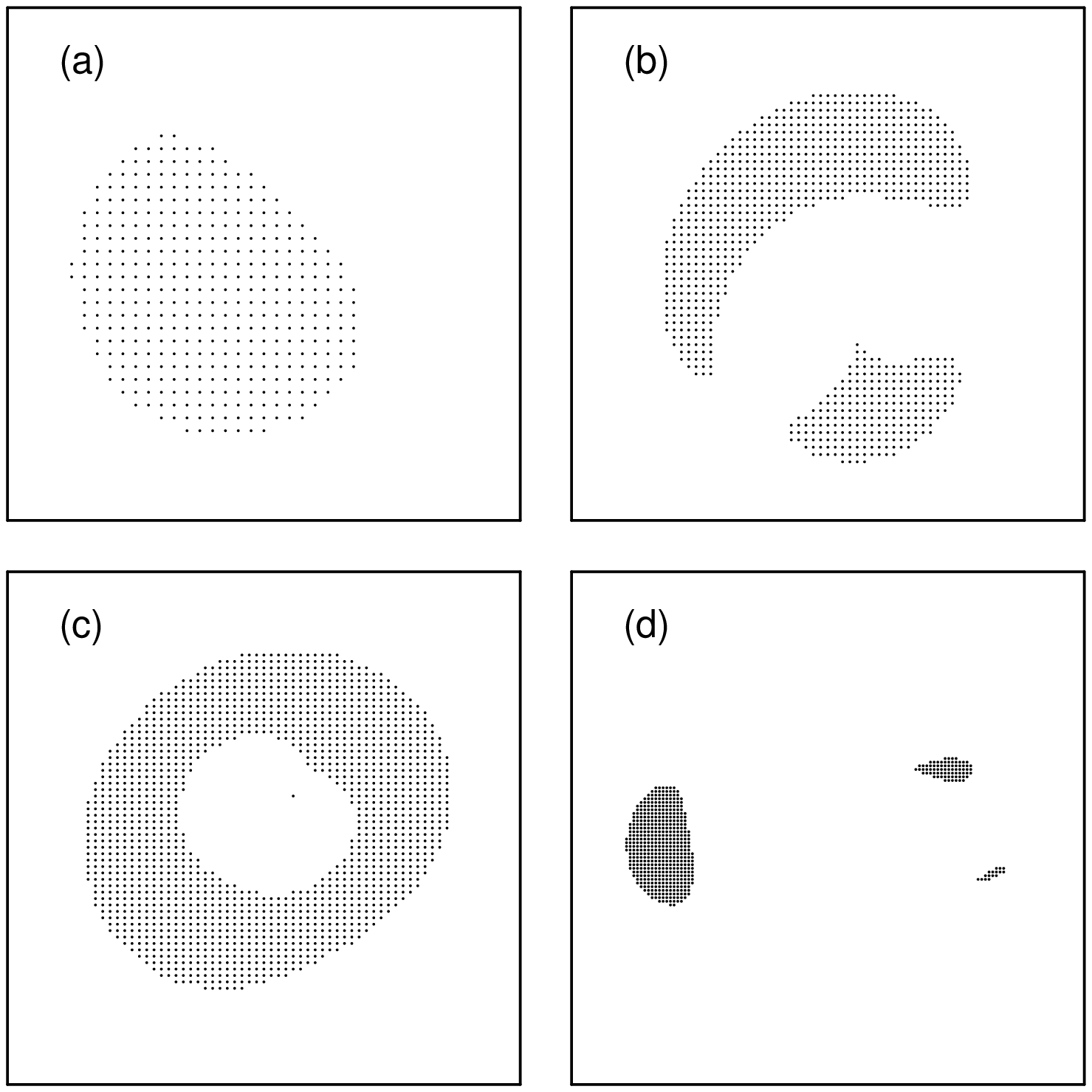}
\vskip-5cm
\line{\hss Fig. 2. Images of distant circular sources \hss}

\bigskip\bigskip

\line{\hss Table 2. Multiple-image Probabilities \hss}
\bigskip

\line{\hss
\vbox{
\halign{
\strut\hfil#\hfil\quad&\hfil#\hfil\quad&\hfil#\hfil\quad
&\hfil#\hfil\quad&\hfil#\hfil\quad&\hfil#\hfil\quad&\hfil#\hfil\cr
\noalign{\hrule}
$\Omega_0$ & $\lambda_0$ & $H_0$ & $\sigma_8$ & \# or runs & Prob \cr
\noalign{\hrule}
1.0 & 0.0 & 65.0 & 1.0 & 37 & 0.040 \cr
0.2 & 0.0 & 65.0 & 0.5 & 38 & 0.088 \cr  
0.5 & 0.0 & 65.0 & 1.0 & 30 & 0.054 \cr
0.5 & 0.0 & 75.0 & 1.0 & 20 & 0.070 \cr
0.7 & 0.0 & 65.0 & 1.1 & 18 & 0.059 \cr
0.5 & 0.5 & 65.0 & 1.0 & 20 & 0.089 \cr
0.2 & 0.8 & 65.0 & 0.8 & 39 & 0.130 \cr
\noalign{\hrule}
}
}\hss}

\bigskip\smallskip

\noindent{\bf References}

\medskip

\noindent 
Bunn, E. F., \& White, M. 1997, {\sl Ap.J.}, {\bf 480}, 6

\noindent
Dressler, A. 1980, {\sl Ap.J.,}, {\bf 236}, 351

\noindent Jarosz\'ynski, M. 1992, {\sl M.N.R.A.S.}, {\bf 255}, 655 

\noindent Premadi, P. 1996, Ph.D. Thesis, University of Texas at Austin

\noindent
Premadi, P., Martel, H., \& Matzner, R. (1998), {\sl Ap.J.}, 
{\bf 493}, 10

\vfill\eject

\end